\documentclass[journal]{IEEEtran}

\usepackage[utf8]{inputenc}   
\usepackage[T1]{fontenc}
\usepackage{amsmath,amssymb}
\usepackage{enumitem}         
\usepackage[pdftex]{graphicx} 
\usepackage{array}
\usepackage[dvipsnames,table,xcolor]{xcolor}
\usepackage{url}              

\definecolor{logging-blue}{RGB}{1, 130, 172}
\definecolor{dark-red}{rgb}{0.3,0.1,0.1}
\definecolor{dark-green}{rgb}{0.1,0.3,0.1}
\definecolor{dark-blue}{rgb}{0.1,0.1,0.5}
\usepackage[colorlinks,allcolors=black]{hyperref} 

\usepackage[nameinlink]{cleveref}
\usepackage[nocompress]{cite} 
\usepackage{microtype}

\usepackage{tikz}
\usetikzlibrary{arrows,automata,arrows.meta}
\usepackage{comment}
\usepackage{float}
\definecolor{anotherRed}{RGB}{244,195,184}
\definecolor{phase5}{HTML}{EBFCFB}

\usepackage[newfloat,frozencache,cachedir=./]{minted}
\usepackage{booktabs}
\usepackage{lipsum}

\renewcommand*{\arraystretch}{1.2}
\newcolumntype{L}[1]{>{\raggedright\let\newline\\\arraybackslash\hspace{0pt}}m{#1}}
\newcolumntype{R}[1]{>{\raggedleft\let\newline\\\arraybackslash\hspace{0pt}}m{#1}}

\usepackage{pgfplots}
\usepgfplotslibrary{colorbrewer}
\usetikzlibrary{pgfplots.statistics, pgfplots.colorbrewer}
\usepackage{pgfplotstable}
\pgfplotsset{compat=1.8}

\usepackage{csquotes}           
\allowdisplaybreaks             

\newcommand\copyrighttext{%
  \tiny \textcopyright 2022 IEEE. Personal use of this material is permitted. Permission from IEEE must be obtained for all other uses, in any current or future media, including reprinting/republishing this material for advertising or promotional purposes, creating new collective works, for resale or redistribution to servers or lists, or reuse of any copyrighted component of this work in other works. Cite this article as follows: J. Vykopal, P. Seda, V. Švábenský, and P. Čeleda. \textit{Smart Environment for Adaptive Learning of Cybersecurity Skills}, in IEEE Transactions on Learning Technologies (Volume: 16, Issue: 3, 01 June 2023). DOI: \href{https://doi.org/10.1109/TLT.2022.3216345}{10.1109/TLT.2022.3216345}.}
\newcommand\copyrightnotice{%
\begin{tikzpicture}[remember picture,overlay]
\node[anchor=south,yshift=12pt] at (current page.south) {\fbox{\parbox{\dimexpr\textwidth-\fboxsep-\fboxrule\relax}{\copyrighttext}}};
\end{tikzpicture}%
}

\begin{document}

\title{Smart Environment for Adaptive Learning of~Cybersecurity Skills}

\author{Jan Vykopal,
        Pavel Seda,
        Valdemar Švábenský,
        and Pavel Čeleda
\thanks{This research was supported by the ERDF project CyberSecurity, CyberCrime and Critical Information Infrastructures Center of Excellence (No. CZ.02.1.01/0.0/0.0/16\_019/0000822).}
\thanks{The authors are affiliated with Masaryk University, Czech Republic. E-mail addresses: \texttt{\{vykopal | celeda\}@ics.muni.cz}, \texttt{seda@fi.muni.cz}, and \texttt{valdemar@mail.muni.cz}.}
\thanks{Manuscript received December 15, 2021; revised August 31, 2022.}}

\markboth{IEEE Transactions on Learning Technologies,~Vol.~16, No.~3, June~2023}%
{Vykopal, Seda, Švábenský, Čeleda: Smart Environment for Adaptive Learning of Cybersecurity Skills}

\maketitle
\copyrightnotice 

\begin{abstract}
Hands-on computing education requires a realistic learning environment that enables students to gain and deepen their skills.
Available learning environments, including virtual and physical labs, provide students with real-world computer systems but rarely adapt the learning environment to individual students of various proficiency and background. 
We designed a unique and novel smart environment for adaptive training of cybersecurity skills. The environment collects a variety of student data to assign a suitable learning path through the training. To enable such adaptiveness, we proposed, developed, and deployed a new tutor model and a training format. 
We evaluated the learning environment using two different adaptive trainings attended by 114 students of various proficiency. The results show students were assigned tasks with a more appropriate difficulty, which enabled them to successfully complete the training. Students reported that they enjoyed the training, felt the training difficulty was appropriately designed, and would attend more training sessions like these. 
Instructors can use the environment for teaching any topic involving real-world computer networks and systems because it is not tailored to particular training.
We freely released the software along with exemplary training so that other instructors can adopt the innovations in their teaching practice.
\end{abstract}

\begin{IEEEkeywords}
Adaptive and intelligent educational systems, intelligent tutoring systems, learning environments, virtual labs, security 
\end{IEEEkeywords}

\IEEEpeerreviewmaketitle

\section{Introduction}
\label{sec:intro}

Mastering cybersecurity requires extensive knowledge and skills, ranging from a wide area of theoretical concepts to practical skills with operating systems, command-line tools, and system vulnerabilities~\cite{mouheb2019cybersecurity}. At the same time, more and more students with different backgrounds are entering the field of cybersecurity~\cite{bashir2017profiling}. As~a~result, it is difficult for instructors to conduct hands-on cybersecurity training that would match the proficiency of all students. 

Existing cybersecurity training offerings are based on static scenarios with limited or no adaptiveness to an individual student~\cite{braghin2020modeldriven}. Although the instructor can intervene to help students interactively, this is feasible only in relatively small classes, and not every student actively asks for help. The interactive help is especially complicated during online training (e.g., forced by restrictions caused by the COVID-19 pandemic~\cite{putri2020impact}).

We see the opportunity to address the instructors' problem and improve the students' learning experience using a \emph{smart learning environment} (SLE). This environment considers students' proficiency and adapts the learning content using data about student actions and performance in ongoing training. 
As a consequence, low-performing students are not overwhelmed by too difficult tasks, and high performers are not bored by too simple assignments. In the end, each student benefits from the adaptive training compared to the static assignments. Instructors benefit from efficient management, as well as monitoring of the learning environment and actions of individual students. An SLE thus saves the precious time of instructors, which they can spend on assisting individual students who struggle.

We reviewed the literature on SLE and related technologies such as remote labs, intelligent tutoring systems, and adaptive learning systems.
There are many works and systems for various learning domains such as engineering, technology, science, foreign languages and mathematics~\cite{Tabuenca2021}. 
However, we have not found any smart network lab that would assign hands-on cybersecurity tasks to students based on their proficiency and performance in ongoing training featuring computer and network systems. Therefore, we have been iteratively developing and evaluating a learning environment with this capability. Since cybersecurity is a complex domain encompassing diverse technical knowledge and skills, creating an SLE for it represents a substantial research challenge.

The aims of this paper are to \emph{i)} introduce the design of a \emph{smart network lab} for training that involves computer networks, operating systems, and vulnerable applications, and \emph{ii)} evaluate the lab in authentic teaching of cybersecurity skills. 
Our smart lab uses an unique tutor model and a training format, which are not present in state-of-the-art network lab environments. 
We evaluated our lab in field studies with 114 students of various proficiency participating in either on-site or remote training sessions. The objectives of the evaluation are to investigate \emph{i)} how efficiently were individual learners distributed to tasks of various difficulty and \emph{ii)} stakeholders' experience of using our lab. The results show that students persisted in the adaptive training and successfully completed more tasks compared to non-adaptive training. The students also reported they enjoyed the adaptive training, felt the training difficulty was appropriate, and would attend more adaptive training sessions.

This paper is organized into seven sections. \Cref{sec:related-work} summarized related work, introduces smart learning environments, their core functions, and existing systems providing these functions for teaching cybersecurity hands-on. 
\Cref{sec:environment} introduces our smart lab for learning cybersecurity skills, used methods, and technological components. 
\Cref{sec:lab-view} details the instructor's and student's view of the SLE.
\Cref{sec:evaluation} describes a case study of using the developed SLE in authentic teaching in on-site and remote settings, and \Cref{sec:results} reports and discusses the results. 
Finally, \Cref{sec:conclusion} 
summarizes our contributions.
\section{Background and Related Work}
\label{sec:related-work}
Our work is related to remote labs, intelligent tutoring systems (ITS) and adaptive learning, and especially to smart learning environments.

Remote labs have been researched, developed, and used for teaching of various science and engineering disciplines for more than two decades~\cite{Ma2006, Alkhaldi2016, Grout2017}. Some labs collect data about students' interaction with the lab to provide learning analytics for teachers and learners~\cite{Wuttke2015, Tulha2022, Orduna2014, GarciaZubia2019}, for instance, an identification of common students’ mistakes and remedial actions\cite{Considine2018, Considine2021}. Other labs provide automated student assessment or personalized assignments for each student~\cite{Benattia2019, Goncalves2018}. However, there is no published lab that would provide adaptive learning features described in this paper.

Research of ITS and adaptive learning environment is well-established~\cite{mousavinasab2021, aleven2016instruction}. There are  examples of successful tutoring systems for various fields of computer science, such as SQL-Tutor~\cite{Mitrovic2016} or ProTuS~\cite{Vesin2018}, or systems created by various authoring tools~\cite{Dermeval2018}, even by non-programmers~\cite{Aleven2016example}. However, to the best of our knowledge, there are no ITS for hands-on cybersecurity training in a networked lab environment.

\subsection{Smart Learning Environments}
A recent and thorough literature review by Tabuenca et al.~\cite{Tabuenca2021} has shown that the term \emph{Smart Learning Environment} is used inconsistently in the literature. The authors consolidated the terminology and synthesized core functions and characteristics of SLEs. In the rest of this paper, we use the terms presented in the review. Its authors concluded that \enquote{the smartness in SLEs is the quality of a system to provide assistance for students or teachers considering their barriers for learning.} 

Next, the review identified four key components of SLEs: 
\begin{enumerate}
    \item \emph{Stakeholders} -- students and teachers.
    \item \emph{Space} -- physical or virtual environment where learning occurs. 
    \item \emph{System} providing smartness to the SLE by its core functions \emph{sense}, \emph{analyze}, and \emph{react}. 
    \item \emph{Tools and technology} that facilitate students learning.    
\end{enumerate}
The \emph{system} collects data from the learning context (the \emph{sense} function), processes the collected data (the \emph{analyze} function), and suggests actions to ease learning constraints (the \emph{react} function). These functions are performed using \emph{tools and technologies} such as data processing or visualization.

Tabuenca et al.~\cite{Tabuenca2021} also identified affordances of SLEs reported in 68 empirical studies published from 2000 to 2019. Here we list the four most frequent affordances.
\begin{enumerate}
    \item \emph{Adaptation, customization, and personalization} (\emph{adaptable} onwards) -- refers to adjusting the learning environment considering the stakeholders’ context, for instance, providing adapted and personalized environment for each student.
    \item \emph{Tracking and monitoring} (\emph{traceable} onwards) -- recording data from the stakeholders’ context throughout learning activities using sensors installed in the environment.
    \item \emph{Feedback and recommendations} (\emph{recommendation} onwards) -- information provided by the SLE based on stakeholders’ actions during learning activities, for instance, providing feedback just after answering the question.
    \item \emph{Patterns, activity, and behavior identification} (\emph{pattern recognition} onwards) -- analysis of the collected data and identification of patterns related to stakeholders' behavior and their context, for example, identification of students' engagement when playing an educational game.
\end{enumerate}

\subsection{Environments for Learning Cybersecurity Skills}

\begin{figure}[!ht]
    \centering
    \begin{tikzpicture}[
  ->,                  
  >=stealth',          
  shorten >=1pt,    
  auto,
  node distance=1.575cm,  
  thick,
  scale=1.0,
  every node/.style={scale=0.9}, 
  every node/.style={scale=0.8}, 
  font=\sffamily
  ]
  \node[initial,state,color=blue,minimum size=1.0cm] (C)       {$T_1$};
  \node[state,color=blue,minimum size=1.0cm] (D) [right of=C]       {$T_2$};
  \node[state,color=blue,minimum size=1.0cm] (F) [right of=D]       {\dots};
  \node[state,color=blue,minimum size=1.0cm] (G) [right of=F]       {$T_{n-1}$};
  \node[state,color=blue,minimum size=1.0cm] (H) [right of=G]       {$T_n$};
  \node[state,minimum size=1.0cm] (E) [right of=H]       {End};
  \path 
        (C) edge node {} (D)
        (D) edge node {} (F)
        (F) edge node {} (G)
        (G) edge node {} (H)
        (H) edge node {} (E);
\end{tikzpicture}
    \caption{Linear structure of training consisting of several tasks ($T$).}
    \label{fig:linear}
\end{figure}
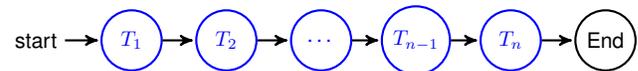

Cybersecurity skills are taught using interactive learning environments featuring emulated networks, IT systems, or applications~\cite{Chouliaras2021,Swann2021}. These learning environments range from relatively simple CTF\footnote{Capture the Flag (CTF) is a popular form of gamified cybersecurity training in an informal setting. A successful solution of a CTF task yields a textual string called \textit{flag}, which the learner submits in the learning environment to prove reaching the solution~\cite{Svabensky2022thesis}.} platforms~\cite{Kucek2020} to sophisticated cyber ranges~\cite{yamin2020}. They enable individual students to learn by solving a set of tasks ($T$), which are often ordered linearly as depicted in \Cref{fig:linear}. 

The completion of each task is assessed by the environment, which checks whether the student submitted the correct answer, generated the expected network traffic, or changed the system state in the required way. Some platforms allow instructors to define static hints, which are provided to students on-demand when needed. Examples of these platforms are Hack The Box~\cite{hackthebox}, TryHackMe~\cite{tryhackme}, Project Ares~\cite{project-ares}, THREAT-ARREST~\cite{hatzivasilis2020modern}, and 
KYPO Cyber Range Platform~\cite{my-2021-FIE-KYPO-CSC}.

The role of the instructors who use these platforms 
shifts from being an active intermediary between learning content and students to a facilitator of learning who employs the platform and its features. 
Once a training starts, the instructors monitor students' progress using the insights automatically provided by the platform, such as those presented in~\cite{oslejsek2021}. The insights are generated using the methods of learning analytics~\cite{handbook-la2017} and educational data mining~\cite{handbook-edm2010}, which leverage data from educational contexts to understand and improve teaching and learning~\cite{romero2020, hundhausen2017}. If the instructors see students who need help, they can intervene appropriately.

\subsection{How Smart Are Existing Environments for Learning Cybersecurity Skills?} \label{sec:sota-smart}

Although Tabuenca et al.~\cite{Tabuenca2021} did not discover any SLE built specifically for learning cybersecurity or related fields such as networking or operating systems, there are a few works that include some of the SLE core functions.

Cyber ranges~\cite{yamin2020} and Capture the Flag platforms~\cite{Kucek2020} are learning technologies for cybersecurity that often employ data collection (the \textit{sense} function). Maennel~\cite{Maennel2020} reviewed digital datasets collected in cybersecurity training, which include timing information, commands, action counts, and input logs. However, as Weiss et al.~\cite{Weiss2016} pointed out, the subsequent analysis of these data (the \textit{analyze} function) is often limited to binary scoring of learners. 

A rare exception is a work by Deng et al.~\cite{Deng2018personalized} who evaluated a personalized lab environment that analyzes student activities. Examples of these activities include \enquote{mouse click, mouse hover, command line activity and time spent inside a virtual machine} for cybersecurity training. Data about these activites are used as features to train a classifier to determine students' learning style. Subsequently, the system personalizes the style and presentation of the study materials for individual students. The SLE proposed by us differs in its goal: we aim to provide learners with adaptively chosen tasks of suitable difficulty.

To conclude, almost no environment for learning cybersecurity skills is advanced enough to offer actionable steps for supporting learning (the \textit{react} function).

\section{Smart Lab for Learning Cybersecurity Skills}
\label{sec:environment}

\begin{figure}[!ht]
    \centering
    \includegraphics[width=1.0\columnwidth]{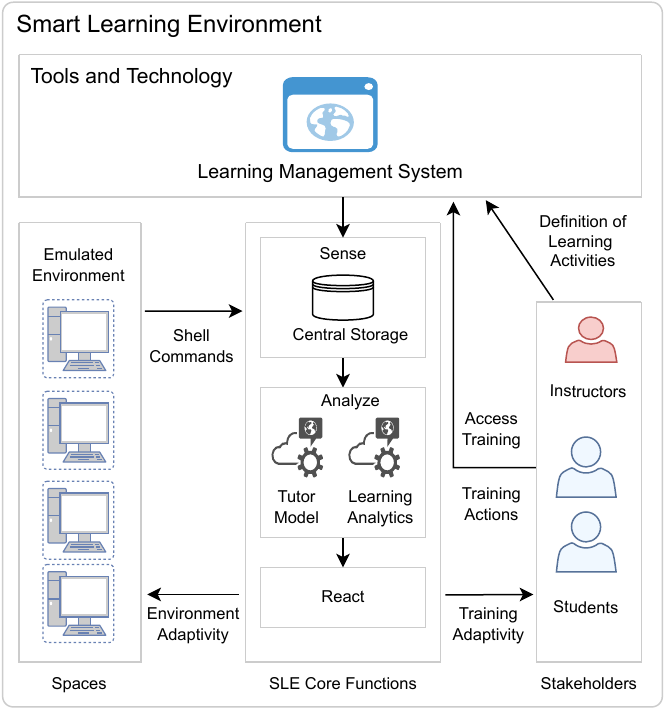}
    \caption{Architecture overview of components of KYPO SLE.}
    \label{fig:dataCollectionArchitectureOverview}
\end{figure}

The proposed smart lab (further \textit{KYPO SLE}) is based on KYPO CRP~\cite{my-2021-FIE-KYPO-CSC}, a platform we have been developing and using for hands-on cybersecurity training. \Cref{fig:dataCollectionArchitectureOverview} shows KYPO SLE mapped to the overall composition of a smart learning environment presented in~\cite[Fig. 3]{Tabuenca2021}.
Here, we detail the key SLE components in the context of learning cybersecurity skills.
\begin{itemize}
    \item \emph{Stakeholders} -- Instructors and students. Instructors prepare and supervise training activities in the virtual learning environment for students who perform these activities.
    \item \emph{Spaces} -- A virtual environment that a student can use from anywhere with a stable Internet connection, most commonly from home, school, or workplace.
    \item \emph{System} -- 
    KYPO CRP enhanced by these SLE core functions:
    \begin{itemize}
        \item \emph{Sense}: Collects actions that students performed in the virtual environment, for instance, commands typed in the emulated environment (training sandboxes) or answers submitted to the training portal (see \Cref{sec:collecitionOfEducationalData}).
        \item \emph{Analyze}: Processes the collected data and provides them as input to a novel \emph{tutor model} described in \Cref{sec:tutorModel}, which determines the most suitable learning path for each student. Also processes the data for creating the visualization of students' progress and performance for both students and instructors.
        \item \emph{React}: Presents the most suitable task for each student based on the output of the tutor model and evaluates the task completion (see \Cref{sec:react-UI}).
        Using the terminology of adaptive learning systems, our SLE provides \emph{task-loop adaptivity}~\cite{aleven2016instruction}.
    \end{itemize}
    \item \emph{Tools and technology} -- The virtual environment students interact with is hosted in a cloud or locally at personal computers (such as a PC in a school lab or students' own laptops). In addition, the SLE is designed so that students need only a web browser to participate in training.  
\end{itemize}

\subsection{Generic Format of an Adaptive Training}\label{sec:adaptive-format}

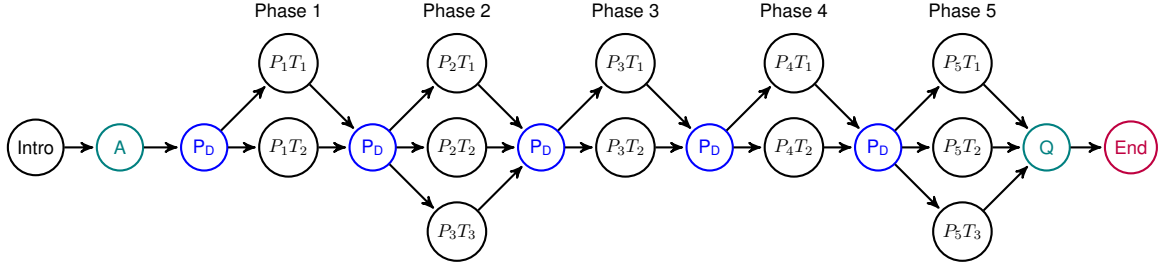
\begin{figure*}[!ht]
    \centering
    \begin{tikzpicture}[
        ->,                  
        >=stealth',          
        shorten >=1pt,    
        auto,
        node distance=1.60cm,  
        thick,
        scale=0.50,
        every node/.style={scale=0.700}, 
        font=\sffamily
    ]
        \node[state,color=black]   (I)                    {Intro};
        \node[state,color=teal] (A) [right of=I] {A};
        
        \node[state,color=blue] (UD1) [right of=A]       {P$_\text{D}$};
        
        \node[state] (U11) [right of=UD1] 		{$P_1 T_2$};
        \node[state] (U12) [above of=U11]       {$P_1 T_1$};
        \node[draw=none] (Unit1) [above of=U11,yshift=1.00cm] {Phase 1};
        
        \node[state,color=blue] (UD2) [right of=U11]       {P$_\text{D}$};
        
        \node[state] (U22) [right of=UD2]       {$P_2 T_2$};
        \node[state] (U21) [above of=U22] 		{$P_2 T_1$};
        \node[state] (U23) [below of=U22]       {$P_3 T_3$};
        \node[draw=none] (Unit2) [above of=U21,yshift=-0.60cm] {Phase 2};
        
        \node[state, color=blue] (UD3) [right of=U22] {P$_\text{D}$};
        
        \node[state] (U32) [right of=UD3]       {$P_3 T_2$};
        \node[state] (U31) [above of=U32] 		{$P_3 T_1$};
        \node[draw=none] (Unit3) [above of=U31,yshift=-0.60cm] {Phase 3};
        
        \node[state, color=blue] (UD4) [right of=U32] {P$_\text{D}$};
        
        \node[state] (U42) [right of=UD4]       {$P_4 T_2$};
        \node[state] (U41) [above of=U42] 		{$P_4 T_1$};
        \node[draw=none] (Unit4) [above of=U41,yshift=-0.60cm] {Phase 4};

        \node[state, color=blue] (UD5) [right of=U42] {P$_\text{D}$};

        \node[state] (U52) [right of=UD5]       {$P_5 T_2$};
        \node[state] (U51) [above of=U52] 		{$P_5 T_1$};
        \node[state] (U53) [below of=U52]       {$P_5 T_3$};
        \node[draw=none] (Unit5) [above of=U51,yshift=-0.60cm] {Phase 5};
        
        \node[state,color=teal] (Q) [right of=U52] {Q};

        \node[state,color=purple] (END) [right of=Q]       {End};
        
        \path (I) edge node {} (A)
        
        (A) edge node {} (UD1)
        
        (UD1) edge node {} (U11)
              edge node {} (U12)
        
        (U11) edge node {} (UD2)
        (U12) edge node {} (UD2)
        
        (UD2) edge node {} (U21)
              edge node {} (U22)
              edge node {} (U23)
        
        (U21) edge node {} (UD3)
        (U22) edge node {} (UD3)
        (U23) edge node {} (UD3)
        
        (UD3) edge node {} (U31)
              edge node {} (U32)
              
        (U31) edge node {} (UD4)
        (U32) edge node {} (UD4)
        
        (UD4) edge node {} (U41)
              edge node {} (U42)

        (U41) edge node {} (UD5)
        (U42) edge node {} (UD5)

        (UD5) edge node {} (U51)
              edge node {} (U52)
              edge node {} (U53)

        (U51) edge node {} (Q)
        (U52) edge node {} (Q)
        (U53) edge node {} (Q)
        
        (Q) edge node {} (END)
        
        ;
    \end{tikzpicture}
    \caption{Graph structure of adaptive cybersecurity training with pre-training assessment (A), decision component (P$_\text{D}$) applying the proposed model, and a~post-training questionnaire (Q). This exemplary training contains five phases ($P_x$) with different number of tasks ($T_y$).}
    \label{fig:adaptive-game-structure-example}
\end{figure*}

To enable the \textit{adaptable} affordance of KYPO SLE, we proposed a generic structure for adaptive cybersecurity training. In general, the training can contain an arbitrary number of phases and tasks. Each phase represents a learning activity. Each task in the phase exercises the same skills but varies in difficulty.
\Cref{fig:adaptive-game-structure-example} shows an example of such structure with five phases: three with two tasks and two with three tasks of various difficulty.

The training consists of several components: the introduction (Intro), the pre-training assessment (A), training phases ($P_x$) including variant tasks ($T_y$), decision components (P$_\text{D}$), and post-training questionnaire (Q).

First, the introduction (Intro) familiarizes the student with the training and communicates necessary information before the training starts. 

The pre-training assessment (A) is the first component of collecting data about students' knowledge and skills. The questions asked in the pre-training assessment are grouped into \emph{question groups} based on their relation to specific training phases. Each question can be assigned into several question groups since they can be relevant to more phases. For each training phase, we set the \textit{minimal ratio} of knowledge to determine whether the student's 
knowledge or self-reported skills are sufficient or not. For example, the minimal ratio can be set to 100\%, which would mean the students need to know answers to all the questions or self-report a defined level of skills for a particular phase. In particular, pre-training assessment should mostly include knowledge quizzes, as students' self-assessment can be inaccurate \cite{vsvabensky2018challenges,mirkovic2014class}. 

The training phases contain tasks ($T_y$) that vary in difficulty but all aim at practicing the same topic. The decision component assigns exactly one task from the given phase. This assignment is based on the student performance in previous phases and on the results of the pre-training assessment. Students interact with their dedicated emulated environment, typically by entering shell commands, to find an answer: proof they completed the task. The student performance is measured by time, used commands, submitted answers, and a solution displayed in the phase. These performance indicators were selected based on the capabilities of the KYPO CRP platform and aligned with the review of metrics in cybersecurity exercises~\cite{Maennel2020}. The tasks are denoted as $T_1$, $T_2$, \dots, $T_n$, where $T_1$ represents the most difficult task in the phase and $T_n$ the easiest task in the same phase. We refer to $T_1$ as the \emph{base task} and $T_2$, \dots, $T_n$ as \emph{variant tasks}. Further, the decision component (P$_\text{D}$) processes the students' performance and knowledge to assign a~suitable task from the training phase.

Finally, the post-training questionnaire (Q) is an optional part of training, which enables instructors to collect immediate feedback from the students. Depending on the training objectives, the post-training questionnaire can be the same or different as the pre-training questionnaire.

\subsection{Sense -- Collect Data}
\label{sec:collecitionOfEducationalData}

KYPO SLE collects answers from the pre-training assessment, training actions, and shell commands from the learning environment. All these data are further required by the tutor model, which selects the most suitable task for each student (see~\Cref{sec:analyze-react}). 

\subsubsection*{Pre-training Assessment and Training Actions}
The Learning Management System (LMS) is a key component of the SLE. It presents students with the pre-training assessment and tasks that have to be completed in the emulated environment.
The LMS collects answers from a questionnaire at the beginning of the training (the state A in \Cref{fig:adaptive-game-structure-example}) and audits training actions that students make while they work on tasks ($P_xT_y$) in the training phases.

The training actions include answers submitted by the student in all phases, the action of revealing the task solution, and the action of correct/wrong answer to complete the task. All these data are timestamped and saved to the central storage. For instance, when a student submits an incorrect answer (e.g., \texttt{.invoices2021}), the system audits current timestamp in Epoch time (e.g., \texttt{1621524941312}), the type of the training action (\texttt{action.training.WrongAnswerSubmitted}), user pseudo-identifier (e.g., \texttt{5}) and the training run identifier (e.g., \texttt{3}). The data are stored as JSON records.

\subsubsection*{Shell Commands}
When students interact with the emulated environment, they enter commands in shells such as BASH or Metasploit Console. These commands are captured at hosts in the environment in real-time and forwarded using the Syslog Protocol~\cite{rfc5424} to the central storage using Elastic Stack~\cite{elk-stack}. The commands are stored in JSON and timestamped with microsecond precision.

For example, 
a command \verb!ssh alice@server! executed by a student in the Linux terminal at a machine in the emulated environment is timestamped and audited using Syslog as a string~(\Cref{fig:command-local}).
Then, it is transformed into JSON and forwarded to the central storage as an entry for further processing \cite{my-2021-FIE-KYPO-CSC}. 
This way, the submitted commands can be correlated with the pre-training assessment and training actions of the same student.

\newcommand{\cmdpart}[2]{\underbrace{\texttt{#1}}_{#2}\ \ \ }
\begin{figure}[ht]
\[
\setlength{\jot}{7pt} 
\footnotesize
\begin{split}
&\cmdpart{Dec 1 2021 15:00:33}{timestamp}
\cmdpart{username=\textquotedbl root\textquotedbl}{username}
\cmdpart{client}{hostname} \\
&\hspace*{6mm}
\cmdpart{src=\textquotedbl 10.10.40.5\textquotedbl}{host\ IP\ address} 
\cmdpart{cmd=\textquotedbl ssh alice@server\textquotedbl}{command}\\
&\hspace*{6mm}
\cmdpart{cmd\_type=\textquotedbl bash\textquotedbl}{command\ type}
\cmdpart{uid=\textquotedbl 1\textquotedbl}{sandbox\ ID}
\cmdpart{wd=\textquotedbl /home\textquotedbl}{working\ directory}
\end{split}
\]
\caption[A log entry for a command executed on one machine in an emulated environment.]{A log entry for a command executed on one machine in an emulated environment \cite{my-2021-FIE-KYPO-CSC}.}
\label{fig:command-local}
\end{figure}

All hosts in the emulated environment use clock synchronization via the network time protocol (NTP)~\cite{rfc5905}. This setting is a key requirement for time-correlating the captured commands with training actions and other data. The architecture for collecting shell commands is detailed in~\cite{svabensky-2021-FIE-logging}.

\subsection{Analyze -- Select the Most Suitable Task} \label{sec:analyze-react}
When designing the \enquote{Analyze} function of KYPO SLE, we had to deal with constraints specific to cybersecurity hands-on training. These include: heterogeneity of training definitions, which can have different phases and relations between them; a limited volume of data to find statistical patterns; complexity of the performed tasks; and the inability to collect more in-depth data about students before the training. We designed a novel tutor model that processes the collected student data and computes the number of the most suitable task in a particular phase for each student~\cite{my-2021-FIE-adaptive}.

\subsubsection*{Tutor model}\label{sec:tutorModel}

Let us denote the variables $\boldsymbol{p}$, $\boldsymbol{k}$, $\boldsymbol{a}$, $\boldsymbol{t}$, and $\boldsymbol{s}$, which are the binary vectors on the correctness or incorrectness of prerequisites for a particular training phase. Vector $\boldsymbol{p}$ is defined as follows:
$\boldsymbol{p} = \begin{pmatrix}p_1 & p_2 & \dots & p_m\end{pmatrix}$, where $m$ is the number of training phases. The other vectors use the analogous notation.
\begin{itemize}
    \item $\boldsymbol{p}$ represents the (in)correctness of answers from the pre-training assessment,
    \item $\boldsymbol{k}$ indicates if the student used the expected key commands in the command line within the given task,
    \item $\boldsymbol{a}$ denotes whether the student submitted the expected answers to the task,
    \item $\boldsymbol{t}$ contains the information if the task was completed in a~predefined time, and
    \item $\boldsymbol{s}$ contains the information whether the student asked to reveal the solution for the task.
\end{itemize} 

The model is defined by the \Crefrange{eq:weightMatrix}{eq:taskAssignment}. 
By \Cref{eq:weightMatrix}, we get the \emph{decision matrix} $\boldsymbol{W}$ with weights for the individual phases' metrics. It is specific for each training phase. The weights represent the relationships between phases and their metrics. The value of the weight determines the importance of the metric to the phase. For instance, consider training with six phases where the third phase deepens the topic exercised in the first phase. In this case, we set the weights in the third matrix so that the selected weights for the metrics from the first phase are non-zero. The other performance metrics with weights set to zero are ignored.

The weights have to be manually set by the instructor since each training is unique. 
The number of decision matrices is equal to the number of training phases. The symbols 
$\pi,\kappa,\alpha,\theta,\sigma$
denote the columns in the decision matrices and the $i=1,\dots, m$ are the rows in the decision matrices.

By \Cref{eq:achievedPerformance} we get the \emph{student's performance} based on the defined metrics and their weights for completed phases. The value of the performance is in the interval of $[0,1]$. In~\Cref{eq:achievedPerformance}, $s$ is multiplied by $a$, $k$, and $t$ to distinguish between students who satisfy $a$, $k$, and $t$ metrics without using a solution and solved the task on their own.

By \Cref{eq:taskAssignment} we get \emph{the number of the most suitable task $y$} in phase $x$ for a particular student (1 is $T_1$, 2 is $T_2$, and so on).

\begin{align}\label{eq:weightMatrix} 
\boldsymbol{W}^{(x)} = \left(w^{(x)}_{ij}\right), i=1,\dots,m,\:\:\: j=\pi,\kappa,\alpha,\theta,\sigma
\end{align}

\begin{equation}\label{eq:achievedPerformance}  \footnotesize
        f(x)=\displaystyle\frac{\sum\limits_{i=1}^{x} \left[p_{i}w_{i\pi}^{(x)} + s_{i} \left(k_{i}w_{i\kappa}^{(x)} + a_{i}w_{i\alpha}^{(x)} +   t_{i}w_{i\theta}^{(x)} + w_{i\sigma}^{(x)}\right)\right]}
        {\sum\limits_{i=1}^{x} \left(w_{i\pi}^{(x)}  + w_{i\kappa}^{(x)} + w_{i\alpha}^{(x)} + w_{i\theta}^{(x)} + w_{i\sigma}^{(x)}\right)}
\end{equation}

\begin{align}\label{eq:taskAssignment} 
    T_{y} = 
    \begin{cases} 
    n_x, \qquad\qquad\qquad\qquad\qquad\qquad \text{if } f(x) \text{ is equal to } 0\\
    \text{trunc}(n_x [1-f(x)])+1, \:\:\,\,\qquad \text{otherwise}
    \end{cases}
\end{align}


where:
\begin{align*}
    x= & \text{ the phase a student is entering},\\
    y= & \text{ the order of the task in a phase},\\
    T_{y}= & \text{ the most suitable task of the phase $x$ for the student}, \\
    n_x= & \text{ the number of variant tasks in the phase $x$}, \\
    p_i= & 
    \begin{cases} 
        1, \: \text{if}\: \text{question group $i$ from A is correctly answered}\\
        0, \: \text{otherwise},
    \end{cases} \\
    k_i= & \text{ commands corresponding to the phase $i$ were used}, \\
    e_i= & \text{ expected time to complete of the phase $i$}, \\
    o_i= & \text{ student's completion time in the phase $i$}, \\
    t_i= &
    \begin{cases} 
        1, \: \text{if}\: o_i < e_i \text{ in phase $i$}\\
        0, \: \text{otherwise}, 
    \end{cases}\\
    s_i= &
    \begin{cases}
         1, \: \text{if}\: \text{the solution of the phase $i$ is \emph{not} displayed}\\
         0, \: \text{otherwise}, 
    \end{cases}\\
    a_i= & \text{ answers corresponding to the phase $i$ were submitted}.
\end{align*}

\subsubsection*{Model Assumptions}
The proposed model requires several assumptions that must be met by any SLE that would use it for hands-on cybersecurity training  \cite{my-2021-FIE-adaptive}.

\begin{itemize}
    \item The learning environment has to collect the required data: the pre-training assessment answers $p$, commands typed by the students $k$, the submitted answers $a$, phase completion time $t$, and the action of displaying the solution $s$.
    \item The model expects that some tasks are related; otherwise, it will heavily rely only on the pre-training assessment that may not be sufficient to capture students' proficiency.
    \item The pre-training assessment question groups have to be mapped to the training phases to distinguish the level of knowledge and self-reported skills for a particular phase.
    \item The model assumes that the tasks in the phases are sorted so that the $T_1$ is the most difficult task, $T_2$, \dots, $T_{n-1}$ are gradually easier tasks than $T_1$, and $T_n$ is the easiest task.
\end{itemize}

To ease the unified design and run of the training, we add the following constraints that simplify the model assumptions:
\begin{itemize}
    \item The students' performance in a phase is evaluated in the same way in all tasks.
    \item The observed metrics are binary. Other metrics of students' performance, such as similarity of the submitted answers to the correct ones, are either unavailable or ignored.
\end{itemize}

The model was developed with the aim to reinforce the cybersecurity training with respect to the commonly used performance metrics \cite{Maennel2020}. Nevertheless, it can be applied in any domain collecting such data.


\subsection{React -- Serve the Selected Task}\label{sec:react-UI}
When the student transitions between phases, the P\textsubscript{D} component (see~\Cref{fig:adaptive-game-structure-example}) is applied. This component uses the model described in~\Cref{sec:tutorModel} to assign the most suitable task in the next phase. When the task is assigned to the student, the task content is shown to the student. Each student can receive different task content.

\begin{figure*}[!ht]
    \centering
    \includegraphics[width=0.95\textwidth]{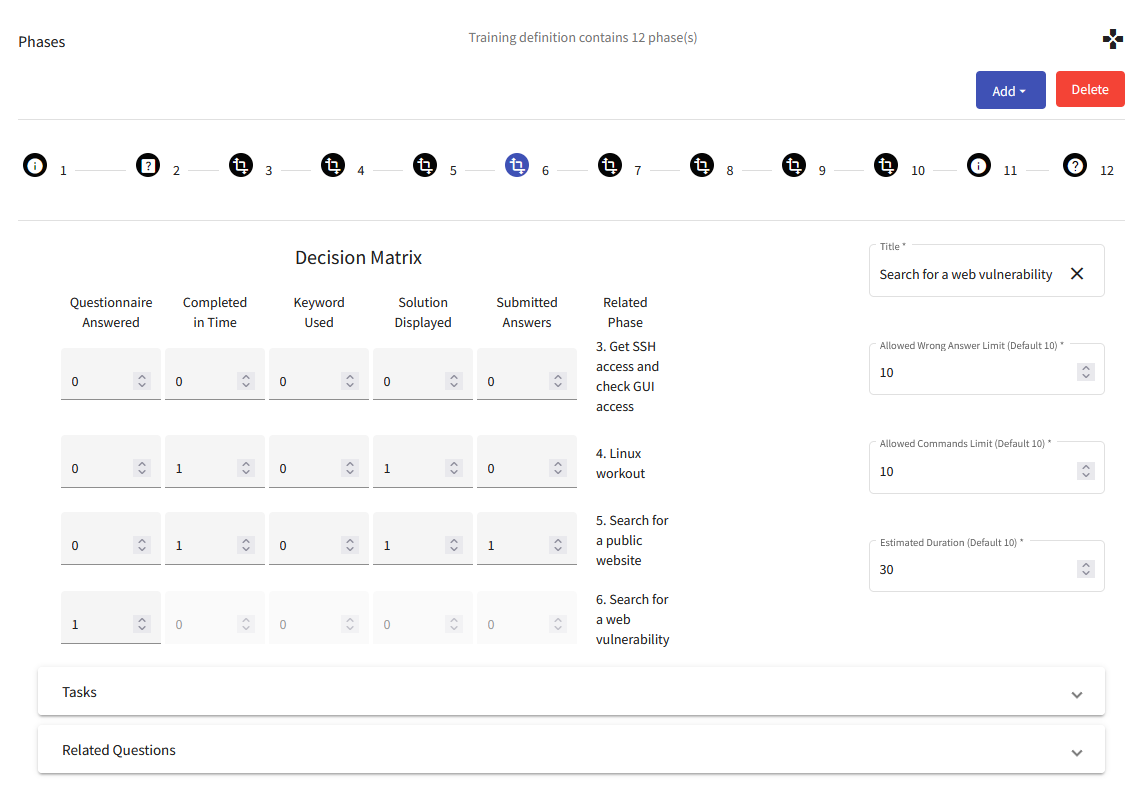}
    \caption{
    Instructor's user interface for setting weights for Phase 6 of the Knowledge Base training via the Decision matrix. The instructor can also set task content and related questions from the pre-training assessment (available under Tasks and Related Questions accordions, respectively). This interface allows setting all weights for all phases in the adaptive training as depicted in 
    \Cref{fig:knowledge-base-relationships}.
    Note: This screenshot from KYPO SLE contains also non-training phases (Intro, A, Q) so the numbering of phases does not align with the phase numbering in other figures (e.g., Sankey diagrams).
    }
    \label{fig:trainingDesignView}
\end{figure*}

\section{Stakeholders' Usage of the Smart Lab} \label{sec:lab-view}

In this section, we describe interactions of instructors and students with KYPO SLE before, during, and after the training.

\subsection{Instructor's View} 

\begin{figure*}[ht]
    \centering
    \includegraphics[width=\textwidth]{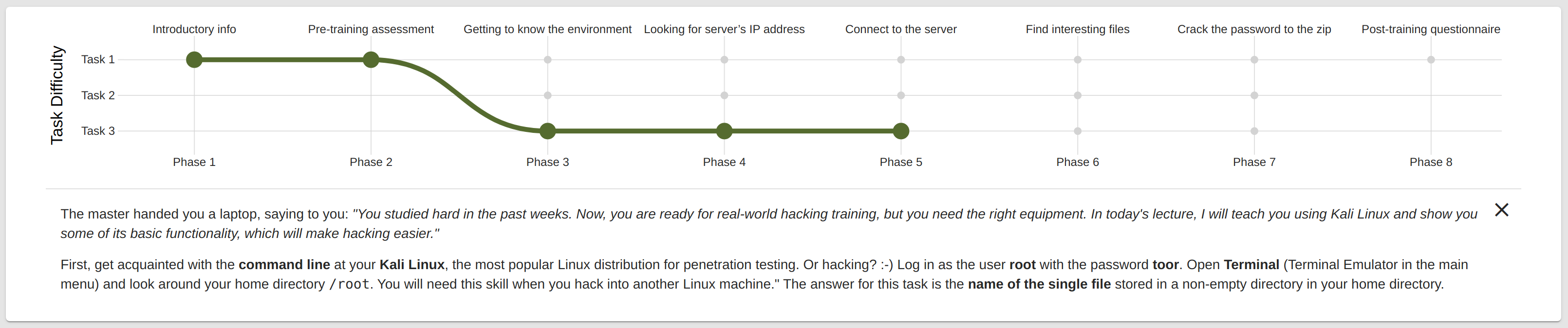}
    \caption{Visualization showing the student's training path and the tasks description.     \label{fig:adaptivePostTrainingVisualization}}
\end{figure*}

\subsubsection{Before the training} 
At first, the instructor(s) have to prepare the training: task assignments, the correct answers, and emulated environment. The learning activities have to be split into several phases as described in~\Cref{sec:adaptive-format}. For each phase, the instructor designs several tasks of varying difficulty to serve students of various proficiency.
Further, for each phase, the instructor sets model weights to define logical relations between phases and their metrics. 

\Cref{fig:trainingDesignView} shows user interface of KYPO SLE for setting the weights of preceding phases for the sixth phase. In this example, the instructor set the weight for \emph{Questionnaire Answered} assigned to the sixth phase, and \emph{Completed in Time} and \emph{Solution Displayed} metrics for the fifth and the fourth phase, and for \emph{Submitted Answers} metric in the fifth phase. The weights set to non-zero values determine which metrics will be used by the SLE for computing the most suitable task in the sixth phase. To ease the design of the model weights, we provide a tool assisting the instructors with the adaptive training design~\cite{seda2022adaptive}.

Finally, the instructor deploys the created training for a particular training session for a predefined number of students. The SLE automatically creates the emulated environment for the defined number of students and generates a unique access token, which the instructor distributes to the students.

\subsubsection{During the training}
After the students enter the training session, the instructor monitors their progress using visual analytics provided by Sankey diagram and a progress chart.

\begin{figure}[ht]
    \centering
    \includegraphics[width=\columnwidth]{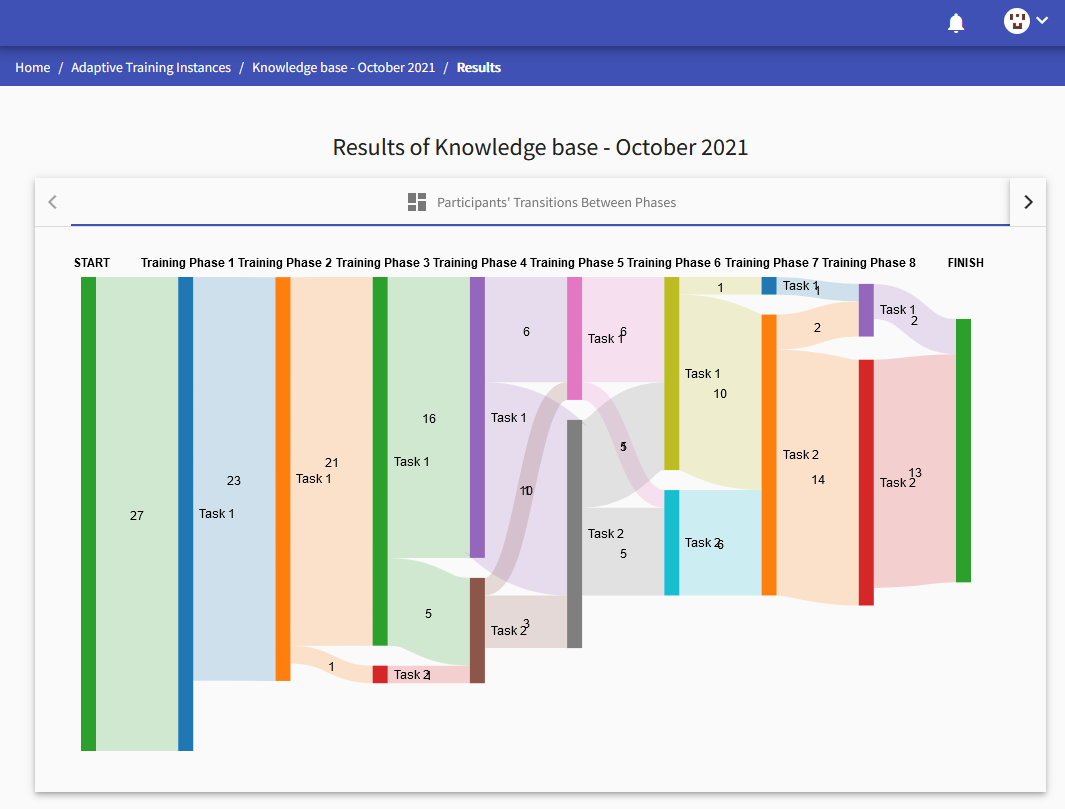}
    \caption{Visualization showing the real-time progress of students during the adaptive training in KYPO CRP. It shows the number of students in particular phases and tasks.}
    \label{fig:instructorsViewDuringTheTraining}
\end{figure}

The Sankey diagram (see~\Cref{fig:instructorsViewDuringTheTraining}) enables the instructor to monitor the overall progress of all students in the training. The instructor might provide additional help to students who enter the easier tasks and still struggle.
The progress chart (see~\Cref{fig:adaptivePostTrainingVisualization}) provides a detailed view of the progress and pathway of a selected student.

\subsubsection{After the training}
When the training is over, the visualization of student progress is shown to the instructor and students. While instructors see the pathways of all students in one view (as in~\Cref{fig:instructorsViewDuringTheTraining}), each student sees only their own pathway (as in~\Cref{fig:adaptivePostTrainingVisualization}). The instructor can easily identify the critical training phases and give feedback to students for future learning or improve the training.

\subsection{Student's View} 
\label{sec:students-view}

\subsubsection{Before the training}
Before the training, the students 
receive a URL to the web portal of KYPO SLE, 
requirements for the student's system used for accessing the SLE, and access token to enter a particular training. Then, the students log into the system using their credentials and enter the access token to start the training. In that moment, one instance of an existing emulated environment is assigned to the student.

\subsubsection{During the training}
First, the students read an introduction to the training and continue with the pre-training assessment of their theoretical knowledge and self-reported levels of skills. After the students complete this assessment, they enter the training phases to exercise cybersecurity skills. The training phases involve practical tasks performed in the student's own instance of the emulated environment. The students are not explicitly informed that training is adapted to their current performance and proficiency.   

\subsubsection{After the training}
When a student finishes the training, their progress is visualized to them to provide feedback and insights for future learning. \Cref{fig:adaptivePostTrainingVisualization} shows an example of such visualization. The student can see their path through the training. If the path moves in the lower parts (variant tasks, such as P3T3), this indicates missing knowledge or skills required by a particular task since the student did not satisfy prerequisites for more difficult tasks (such as P3T2 or P3T1). Additionally, the student can see the assignment of any task by selecting bullets in the grid representing all tasks in the training. 
\section{Case Study Setup} \label{sec:evaluation}

This section describes the case study of using KYPO SLE in teaching practice. The study evaluates the smart features of the learning environment in different contexts.

\subsection{Study Objectives}

The objective of the study is to investigate 
\emph{i)} how efficiently were individual learners distributed to tasks of various difficulty and \emph{ii)} stakeholders' experience of using KYPO SLE.
In the case of students, we are interested whether the lab eases their learning. 
In particular, we study whether low-performing students are provided with easier tasks, which enables them to complete the training in expected time.
In the case of instructors, we analyze how much time and effort is saved by KYPO SLE compared to a manual assignment of training tasks to each student by instructors.
Our study is conducted in two different contexts: a training session with and without the instructor's supervision.

\begin{figure*}[!ht]
    \centering
    \includegraphics[width=0.8\linewidth]{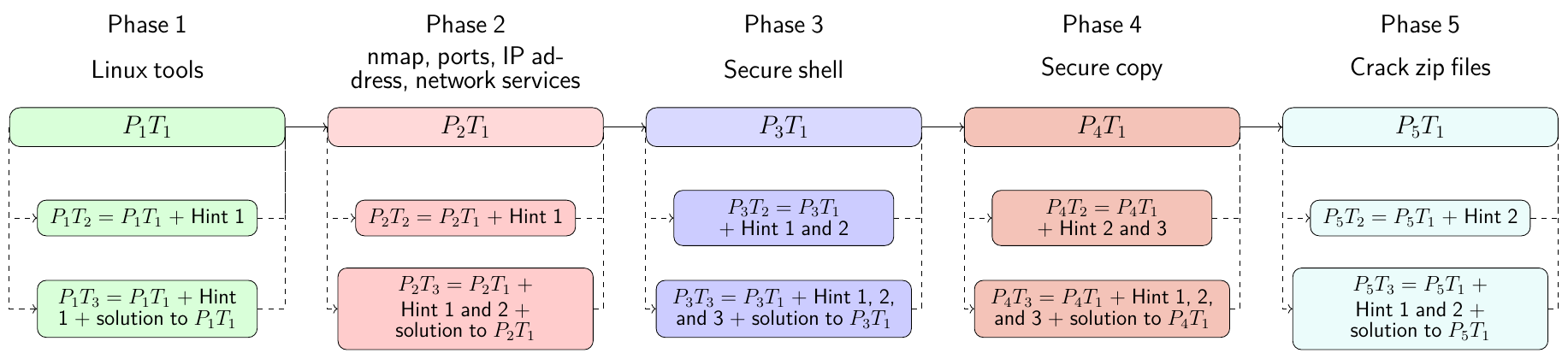}
    \caption[Phases of the Junior Hacker adaptive training. 
    Assignments of variant tasks enhance base tasks by hints or the solution.]{Phases of the Junior Hacker adaptive training. 
    Assignments of variant tasks enhance base tasks by hints or the solution~\cite{my-2021-FIE-adaptive}.}
    \label{fig:junior_hacker_phases}

    \bigskip

    \includegraphics[width=\linewidth]{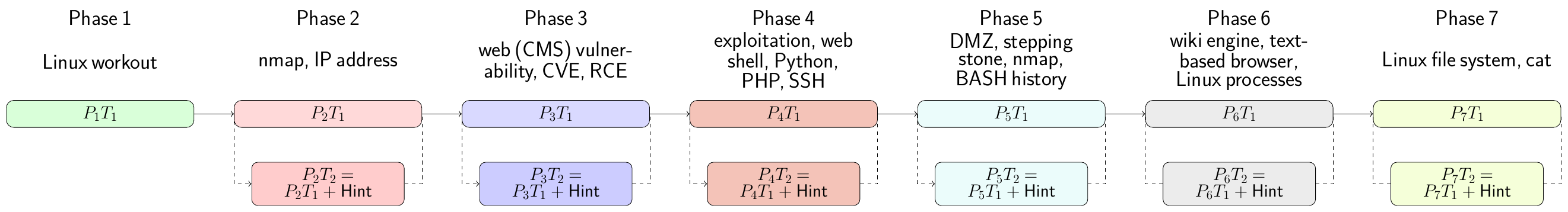}
    \caption{Phases of the Knowledge Base adaptive training that follows the proposed generic format. The assignment of Task 2 enhances Task 1 by a hint.}
    \label{fig:knowledge_base_phases}
\end{figure*}

\subsection{Study Design} 

We followed the approach of action research~\cite{Avison1999}, which is closely related to design-based research~\cite{Anderson2012}. Both methods are extensively used in applied and educational research. Their methodology involves developing a prototype that addresses a practical problem, testing it in an authentic context, performing a small-scale evaluation, and iterating the development further based on the lessons learned from the evaluation~\cite{Svabensky2022thesis}.

At first, we enhanced our existing KYPO Cyber Range Platform with data collection features described in \Cref{sec:collecitionOfEducationalData} and implemented a prototype of the tutor model presented in \Cref{sec:tutorModel}. Along with that, we created the first adaptive training following the proposed generic format described in \Cref{sec:adaptive-format}. Then, we held the first training session with 24 participants and published the initial results~\cite{my-2021-FIE-adaptive}.

Based on the lessons learned, we integrated the prototype of the tutor model with a user interface described in \Cref{sec:react-UI} and created a full-fledged SLE, which is publicly available~\cite{kypo-website}. We then designed another adaptive training and held additional training sessions to show the versatility of the training format and KYPO SLE. 
In total, we held ten training sessions with 114 participants in two different trainings.  

Both trainings were designed to last two hours to fit our classes. They were first tested by experienced instructors and then used in this study. The second training was intentionally designed with more phases but less tasks to highlight capabilities and limitations of the proposed training format and tutor model. 

\begin{figure}[!ht]
    \centering
    \includegraphics[width=\columnwidth]{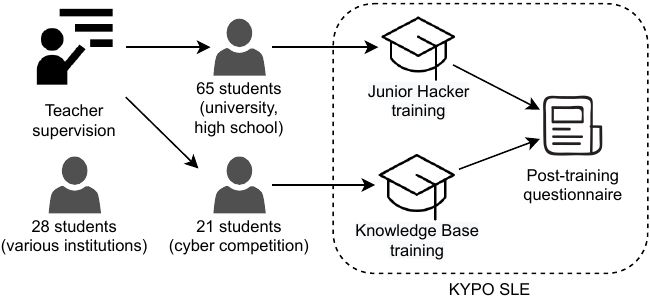}
    \caption{Study design: 114 students completed one of two adaptive trainings deployed in KYPO SLE and answered a post-training questionnaire. Most training sessions (86 students) were facilitated by the instructor, but some (28 students) were not.}
    \label{fig:study-overview}
\end{figure}

\Cref{fig:study-overview} visualizes the study framework. The role of instructors during the supervised sessions was only to provide technical assistance related to using the SLE. Specifically, the instructors did not provide any hints on training tasks.

\subsection{Adaptive Trainings} \label{sec:trainings}

\subsubsection*{Junior Hacker Training} 

\begin{figure}[!ht]
    \centering
    \begin{tikzpicture}[
        ->,                  
        >=stealth',          
        shorten >=1pt,    
        auto,
        node distance=2.85cm,  
        thick,
        scale=0.99,
        every node/.style={scale=0.70}, 
        font=\sffamily
    ]
        \node[state,fill=green!15] (P1)        {$P_1$};
        \node[above of=P1,yshift=-2.115cm] (P1Weight)        {$w_{1\pi}^{(1)}$};
        \node[state,fill=red!15] (P2) [right of=P1]      {$P_2$};
        \node[above of=P2,yshift=-2.115cm] (P2Weight)        {$w_{2\pi}^{(2)}$};
        \node[state,fill=blue!15] (P3) [right of=P2]      {$P_3$};
        \node[above of=P3,yshift=-2.115cm] (P3Weight)        {$w_{3\pi}^{(3)}$};
        \node[state,fill=anotherRed] (P4) [right of=P3]      {$P_4$};
        \node[above of=P4,yshift=-2.115cm] (P4Weight)        {$w_{4\pi}^{(4)}$};
        \node[state,fill=phase5] (P5) [right of=P4]      {$P_5$};
        \node[above of=P5,yshift=-2.115cm] (P5Weight)        {$w_{5\pi}^{(5)}$};
        
        \path 
        (P1) edge[dotted] node {} (P2)
        (P2) edge[dotted] node {} (P3)
        (P3) edge[dotted] node {} (P4)
        (P4) edge[dotted] node {} (P5)
        ;
        
        \path (P2) edge [bend left=30] node[xshift=0.55cm] {$w_{1\kappa}^{(2)},w_{1\theta}^{(2)},w_{1\sigma}^{(2)}$} (P1);
        
        \path (P3) edge [bend left=30] node[xshift=0.2cm] {$w_{2\sigma}^{(3)}$} (P2);
        \path (P3) edge [bend right=35] node[yshift=0.75cm,xshift=0.1cm] {$w_{1\kappa}^{(3)},w_{1\theta}^{(3)},w_{1\sigma}^{(3)}$} (P1);
    
        \path (P4) edge [bend right=50] node[yshift=0.75cm,xshift=0.1cm] {$w_{3\alpha}^{(4)},w_{3\theta}^{(4)},w_{3\sigma}^{(4)}$} (P3);
        \path (P4) edge [bend left=45] node[xshift=-0.35cm] {$w_{2\theta}^{(4)},w_{2\sigma}^{(4)}$} (P2);
        \path (P4) edge [bend left=53] node {$w_{1\theta}^{(4)},w_{1\sigma}^{(4)}$} (P1);
        
        \path (P5) edge [bend left] node[xshift=-0.3cm,yshift=0.1cm] {$w_{4\sigma}^{(5)}$} (P4);
        \path (P5) edge [bend left=45] node {$w_{3\sigma}^{(5)}$} (P3);
        \path (P5) edge [bend right=40] node[yshift=0.75cm] {$w_{2\sigma}^{(5)}$} (P2);
        \path (P5) edge [bend right=50] node[yshift=0.70cm] {$w_{1\kappa}^{(5)},w_{1\alpha}^{(5)},w_{1\theta}^{(5)},w_{1\sigma}^{(5)}$} (P1);
    \end{tikzpicture}
    \caption[The relationships between all phases of Junior Hacker training. $P_x$ is a phase $x$ and $w_{ij}^{(x)}$ is weight for phase $x$ and metric $ij$]{The relationships between all phases of Junior Hacker training. $P_x$ is a phase $x$ and $w_{ij}^{(x)}$ is weight for phase $x$ and metric $ij$ \cite{my-2021-FIE-adaptive}.}
    \label{fig:adaptive-game-relationships}
\end{figure}
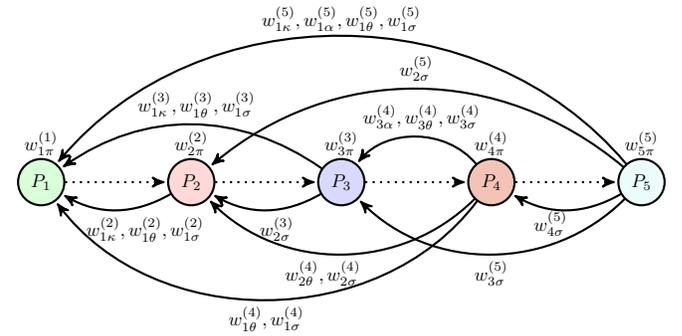

This training consists of the pre-training assessment with eight questions and five phases covering topics depicted in~\Cref{fig:junior_hacker_phases}. Each training phase features one base task and two variant tasks, including one presenting the step-by-step solution. The task with the solution is assigned to students who would not match any phase prerequisites. In the first training phase, basic Linux tools are practiced in three tasks ($P_1T_1$, $P_1T_2$, and $P_1T_3$). Task $P_1T_2$ contains the same assignment as $P_1T_1$ and provides Hint~1. The third task $P_1T_3$ contains the assignment from $P_1T_1$ with Hint~1 and the solution to that task. The subsequent training phases apply the same pattern that differs only in the content of the tasks, hints, and solution provided. 
The relationships between the training phases expressed as weights of each phase in the proposed tutor model are shown in~\Cref{fig:adaptive-game-relationships}.

\subsubsection*{Knowledge Base Training} 

\begin{figure}[!ht]
    \centering
    \begin{tikzpicture}[
        ->,                  
        >=stealth',          
        shorten >=1pt,    
        auto,
        node distance=2.10cm,  
        thick,
        scale=0.5,
        every node/.style={scale=0.66}, 
        font=\sffamily
    ]
        \node[state,fill=green!15] (P1)        {$P_1$};
        \node[state,fill=red!15] (P2) [right of=P1]      {$P_2$};
        \node[above of=P2,yshift=-1.315cm] (P2Weight)        {$w_{2\pi}^{(2)}$};
        \node[state,fill=blue!15] (P3) [right of=P2]      {$P_3$};
        \node[above of=P3,yshift=-1.315cm] (P3Weight)        {$w_{3\pi}^{(3)}$};
        \node[state,fill=anotherRed] (P4) [right of=P3]      {$P_4$};
        \node[above of=P4,yshift=-1.315cm] (P4Weight)        {$w_{4\pi}^{(4)}$};
        \node[state,fill=phase5] (P5) [right of=P4]      {$P_5$};
        \node[above of=P5,yshift=-1.315cm] (P5Weight)        {$w_{5\pi}^{(5)}$};
        \node[state,fill=gray!15] (P6) [right of=P5]      {$P_6$};
        \node[above of=P6,yshift=-1.315cm] (P6Weight)        {$w_{6\pi}^{(6)}$};
        \node[state,fill=lime!15] (P7) [right of=P6]      {$P_7$};

        \path 
        (P1) edge[dotted] node {} (P2)
        (P2) edge[dotted] node {} (P3)
        (P3) edge[dotted] node {} (P4)
        (P4) edge[dotted] node {} (P5)
        (P5) edge[dotted] node {} (P6)
        (P6) edge[dotted] node {} (P7)
        ;
        
        \path (P3) edge [bend right=45] node[xshift=0.65cm,yshift=0.68cm] {$w_{3\theta}^{(1)},w_{3\sigma}^{(1)}$} (P1);
        \path (P3) edge [bend left=45] node[xshift=-0.1cm,yshift=-0.1cm] {$w_{2\alpha}^{(3)},w_{2\theta}^{(3)},w_{2\sigma}^{(3)}$} (P2);
    
        \path (P4) edge [bend left=45] node[xshift=1.8cm,yshift=0.0cm] {$w_{1\alpha}^{(4)},w_{1\theta}^{(4)},w_{1\sigma}^{(4)}$} (P1);
        \path (P4) edge [bend right=60] 
        node[xshift=0.0cm,yshift=0.85cm] {$w_{3\alpha}^{(4)},w_{3\theta}^{(4)},w_{3\sigma}^{(4)}$} (P3);

        \path (P5) edge [bend right=45] node[xshift=0.8cm,yshift=0.68cm] {$w_{1\alpha}^{(5)},w_{
        1\theta}^{(5)},w_{1\sigma}^{(5)}$} (P1);

        \path (P6) edge [bend left] node[xshift=1.2cm,yshift=+0.2cm] {$w_{4\theta}^{(6)}$} (P4);

        \path (P7) edge [bend right=45] node[yshift=0.685cm] {$w_{1\alpha}^{(7)},w_{1\theta}^{(7)},w_{1\sigma}^{(7)}$} (P1);
        \path (P7) edge [bend left=40] node[yshift=-0.1cm] {{\bf $w_{4\theta}^{(7)}$}} (P4);
        
    \end{tikzpicture}
    \caption{The relationships between all phases of Knowledge Base training. $P_x$ is a phase $x$ and $w_{ij}^{(x)}$ is weight for phase $x$ and metric $ij$.}
    \label{fig:knowledge-base-relationships}
\end{figure}
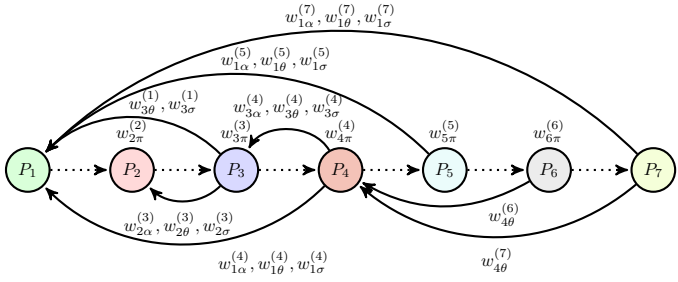

This training consists of the pre-training assessment with eight questions and seven phases covering topics depicted in \Cref{fig:knowledge_base_phases}. Each phase contains one base task and one variant task, which enhances the assignment of the base task with a specific recommended tool or steps needed for finishing the phase. In contrast to the Junior Hacker training, this training contains fewer inter-related phases, as shown in \Cref{fig:knowledge-base-relationships}. However, the student performance in the first phase on Linux essentials ($P_1$) is considered when determining the suitable task in all other phases but the second phase ($P_2$). This was a design decision motivated by i) the fact the basic skills to use the Linux system were a strong prerequisite in this training, and ii) the intent to demonstrate the versatility of the proposed training format and tutor model.

\subsection{Participants}

In total, 114 individuals of diverse demographic characteristics (age, education, experience, and background) participated in our study. The participants' age ranged from 18 to 37. They consisted of high school students, university students, and university graduates, all focusing on computing and related technical disciplines. Since the participants' expertise in cybersecurity varied, they represented a suitable sample for demonstrating the capabilities of our adaptive SLE.

86 participants attended the training under the supervision of one, two, or three instructors, either on-site or remote via video conference. 
65 participants were undergraduate students and graduates of Masaryk University (MU) and the Brno University of Technology (BUT), both located in Brno, Czech Republic. In addition, this group included 4 high school students completing an internship at Masaryk University.
21 participants were senior high school students and bachelor students of other universities, the finalists of the Czech national cybersecurity competition.

In addition, 28 participants attended the training remotely without any guidance (unsupervised training). They came from various institutions including industry companies (such as IBM and Kyndryl) or the two universities (Masaryk University and the Brno University of Technology).

\Cref{table:study-participants} summarizes the information about the trainings. All participants attended voluntarily because of their interest in security. 

\begin{table}[!htb]
\caption{Information about the field studies and the participants.\\MU = Masaryk University, Czech Republic.\\BUT = Brno University of Technology, Czech Republic.}
\footnotesize
\centering
\rowcolors{2}{gray!10}{white}
\renewcommand*{\arraystretch}{1.1}
\begin{tabular}{L{1.65cm}L{1.5cm}L{1.6cm}R{2.25cm}}
\textbf{Training date} & \textbf{Training modality} & \textbf{Participants' institution} & \textbf{Survey responses / num. participants} \\
\hline
Dec 2, 2020 & remote   & MU 
&  9 / 9 \\
Dec 4, 2020 & remote   & MU &  7 / 7 \\
Dec 11, 2020 & remote   & MU &  4 / 4 \\
Jan 14, 2021 & remote   & MU &  4 / 4 \\
May 25, 2021 & remote   & MU &  19 / 19 \\
May 26, 2021 & remote   & BUT &  8 / 10 \\
May 28, 2021 & remote   & BUT & 8 / 8 \\
Jul 22, 2021 & hybrid   & Various &  17 / 21 \\
Sep 9, 2021 & on-site  & High school &  4 / 4 \\
Oct--Nov 2021 & unsupervised  & Various &  15 / 28 \\
\hline
 & & & \textbf{Total: 95 / 114} \\
\end{tabular}
\label{table:study-participants}
\end{table}

\subsection{Data Collection}

The participants were assigned the Junior Hacker or Knowledge Base training described in \Cref{sec:trainings}. They were informed that the estimated time for completing the training is up to two hours. The supervised training sessions were held on-site in a computer lab or remotely via video conference in a time period between December 2020 and September 2021. The primary role of the instructor(s) was only to assist students with access to the virtual lab or to troubleshoot any technical issues that might occur during the training. In contrast, the unsupervised session took place without any instructor's presence and support. Students could choose any time in October and November 2021 when they wanted to take the training and interacted only with our lab. 

We collected all data available in KYPO SLE, i.e., students' answers to questions from the pre-training assessment, training actions, and shell commands. 
Both trainings contain a post-training Likert-scale questionnaire about their training experience (see \Cref{tab:postGameAssessmentQuestions}).
Students who did not finish the training (i.e., did not reach the post-training questionnaire) were asked to fill in an additional questionnaire about issues they encountered during the training.

The study was waived from review by the university institutional review board as the collected data are anonymous and reported aggregately. In addition, all participants provided informed consent to use the collected data for research purposes.

\section{Results and Discussion} \label{sec:results}

\input{figures/boxplot_one_figure}

We now report and discuss the results of the study. We distinguish training sessions with instructor supervision (on-site and remote) and without any supervision (fully remote). Next, we discuss the effort required to run adaptive training with and without the SLE. Finally, we report limitations of the study and lessons learned.

\subsection{Adaptive Training with Instructor's Supervision}

\subsubsection*{Junior Hacker Training} 
\begin{figure}[!ht]
    \centering
    \includegraphics[clip, width=0.95\columnwidth]{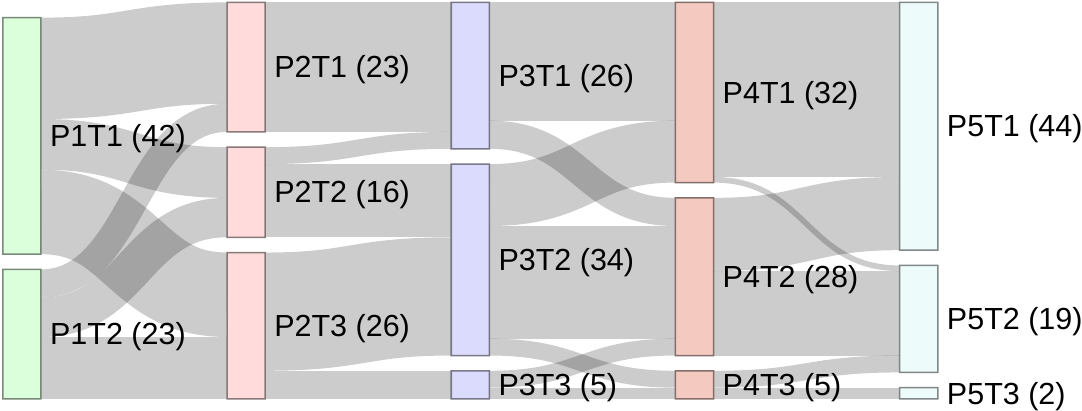}
    \caption{Transitions of 65 students between particular tasks in Junior Hacker training. $P_x T_y$ denotes task $T_y$ in the phase $P_x$. The number of students solving the task is in brackets.}
    \label{fig:phases-flow-JH}

    \bigskip

    \includegraphics[clip, width=0.95\columnwidth]{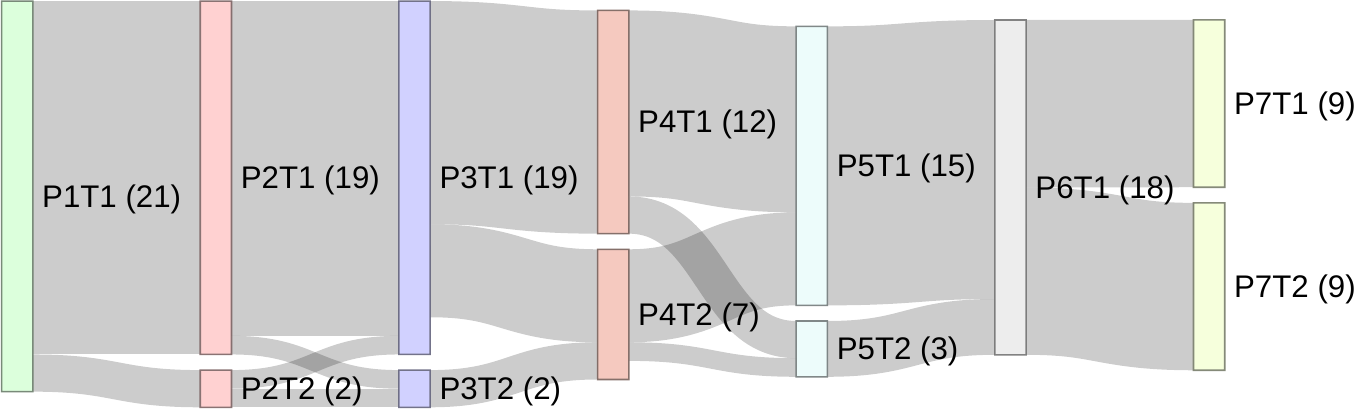}
    \caption{Transitions of 21 students between particular tasks in Knowledge Base training. $P_x T_y$ denotes task $T_y$ in the phase $P_x$. The number of students solving the task is in brackets. The two students quit the training in phase $P_3$.}
    \label{fig:phases-flow-KB}
\end{figure}

This training was finished by all 65 participants.
\Cref{fig:phases-flow-JH} shows the transitions of all participants between tasks ($P_x T_y$) in all training phases of this training. The diversity of transitions shows that the SLE enabled all participants to finish the training, yet by completing less difficult tasks. 

Further, the transitions from more difficult to easier tasks between phases 
indicate that the participants had different issues with different tasks. In the first phase, 23~students assessed their knowledge of Linux basic commands as \enquote{None} or \enquote{Low}. These answers determined the $P_1 T_2$ task for them. In the second phase, $w_{1\kappa}^{(2)},w_{1\theta}^{(2)},w_{1\sigma}^{(2)}$, and $w_{2\pi}^{(2)}$ metrics were evaluated. 23 students were assigned to the hardest (base) task $P_2 T_1$ since they correctly answered the question related to 
Phase~2
and successfully finished 
Phase~1.
16 students were assigned to $P_2T_2$ mostly due to their inability to complete 
Phase~1
in the expected time; others entered too many commands or did not correctly answer the question assigned to 
Phase~2.
The last group of 26 students was assigned to the $P_2 T_3$ task mainly since they claimed to have \enquote{None} or \enquote{Low} skills in searching for opened network ports. In total, 51 students incorrectly answered the question $w_{2\pi}^{(2)}$ assigned to $P_2$, 25 students exceeded the shell commands limit $w_{1\kappa}^{(2)}$ in $P_1$, 29 students exceeded the expected time $w_{1\theta}^{(2)}$ in $P_1$, and six students displayed the solution $w_{1\sigma}^{(2)}$ in $P_1$. In the remaining phases, the students were assigned $T_1$ if they performed well or the other tasks ($T_2$ or $T_3$) due to various issues in related phases or pre-training assessment.

\subsubsection*{Knowledge Base Training}
This training was finished by 18 out of 21 (86\%) participants.
\Cref{fig:phases-flow-KB} shows the transitions of 21 participants between tasks ($P_x T_y$) in the phases of Knowledge Base training. 
This training session was attended by the senior high school students and undergraduates who were finalists of the Czech national cybersecurity competition. 

Although we expected better performance of this group, \Cref{fig:phases-flow-KB} shows that students also solved easier variants of the tasks in all phases except Phase 1. This phase named \enquote{Linux workout} contains only one task, so all the students were assigned to it. In the second phase, two students failed to answer that the \texttt{nmap} tool is used for scanning network ports. In the third phase, two students were provided with the $P_3 T_2$ task. One student revealed solutions in the first two phases, exceeded the estimated time in $P_1$, and failed to answer the questions relevant to the third phase. The other student exceeded the time in the first two phases and failed to answer the question assigned to the third phase. Further, two students exited the training. In the third phase, seven students fell into the $P_4 T_2$ task. Out of the seven students, two revealed the solution from Phase 1 and 3. The other five students had different issues: one submitted too many wrong answers and revealed the solutions, and the others failed to complete the previous phases in an expected time, submitted too many wrong answers, and revealed the solutions. In the fourth, fifth, and sixth phase, the students faced various issues such as exceeding the time to complete, submitting wrong answers, revealing solutions, or providing incorrect answers from pre-training assessment. Due to these deficiencies, the students were assigned easier tasks in the respective phases.


\begin{figure}[!ht]
    \begin{tikzpicture}
      \begin{axis}
        [
            width=0.46\textwidth,
            height=5cm,
            boxplot/draw direction=y,
            every median/.style=very thick,
            xtick={1,2,3,4,5,6,7,8,9,10,11,12},
            xticklabels={,Q1, ,Q2, ,Q3, ,Q4, ,Q5, ,Q6},
            x tick label style={
                text width=2.5cm,
                align=center,
                xshift=-0.6cm,
            },
            ytick={1,2,3,4,5},
            yticklabels={Not at all, Slightly, Moderately, Much, Very much},
        ]
        \addplot[color=purple,
            boxplot prepared={
              median=4,
              upper quartile=4,
              lower quartile=3,
              upper whisker=5,
              lower whisker=1,
              every median/.style=very thick
            },
        ] coordinates {};
        \addplot[color=blue,
            boxplot prepared={
              median=4,
              upper quartile=4,
              lower quartile=3,
              upper whisker=5,
              lower whisker=1,
              every median/.style=very thick
            },
        ] coordinates {};
        \addplot[color=purple,
            boxplot prepared={
              median=3,
              upper quartile=4,
              lower quartile=2,
              upper whisker=5,
              lower whisker=1,
              every median/.style=very thick
            },
        ] coordinates {};
        \addplot[color=blue,
            boxplot prepared={
              median=3,
              upper quartile=4,
              lower quartile=2,
              upper whisker=5,
              lower whisker=1,
              every median/.style=very thick
            },
        ] coordinates {};
        \addplot[color=purple,
            boxplot prepared={
              median=4,
              upper quartile=5,
              lower quartile=3,
              upper whisker=5,
              lower whisker=2,
              every median/.style=very thick
            },
        ] coordinates {};
        \addplot[color=blue,
            boxplot prepared={
              median=3,
              upper quartile=4,
              lower quartile=3,
              upper whisker=5,
              lower whisker=2,
              every median/.style=very thick
            },
        ] coordinates {};
        \addplot[color=purple,
            boxplot prepared={
              median=2,
              upper quartile=3,
              lower quartile=1,
              upper whisker=5,
              lower whisker=1,
              every median/.style=very thick
            },
        ] coordinates {};
        \addplot[color=blue,
            boxplot prepared={
              median=2,
              upper quartile=3,
              lower quartile=1,
              upper whisker=3,
              lower whisker=1,
              every median/.style=very thick
            },
        ] coordinates {};
        \addplot[color=purple,
            boxplot prepared={
              median=3,
              upper quartile=4,
              lower quartile=2,
              upper whisker=5,
              lower whisker=1,
              every median/.style=very thick
            },
        ] coordinates {};
        \addplot[color=blue,
            boxplot prepared={
              median=3,
              upper quartile=3.5,
              lower quartile=2,
              upper whisker=5,
              lower whisker=1,
              every median/.style=very thick
            },
        ] coordinates {};
        \addplot[color=purple,
            boxplot prepared={
              median=5,
              upper quartile=5,
              lower quartile=4,
              upper whisker=5,
              lower whisker=1,
              every median/.style=very thick
            },
        ] coordinates {};
        \addplot[color=blue,
            boxplot prepared={
              median=4,
              upper quartile=4,
              lower quartile=2,
              upper whisker=5,
              lower whisker=1,
              every median/.style=very thick
            },
        ] coordinates {};
    
      \end{axis}
    \end{tikzpicture}
    \caption{Post-training questionnaire answers to Q1--Q6 in the survey from 80 students (red -- Junior Hacker, blue -- Knowledge Base).}
    \label{fig:questionnaireResults}
\end{figure}
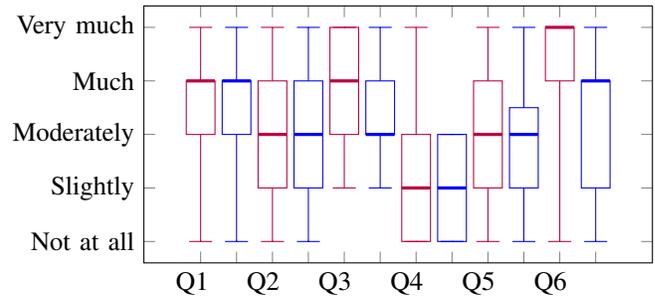
\vspace{-1em}
\begin{table}[!ht]
    \centering
    \caption{Wording of the post-training questionnaire \cite{my-2021-FIE-adaptive}.}
    \label{tab:postGameAssessmentQuestions}
    \small
    \begin{tabular}{p{0.45cm}|p{7.3cm}}
    \hline 
    \textbf{No.} & \textbf{Question} \\  \hline
        Q1 & Did you feel the tasks were designed so that you can complete the training in a timely manner?  \\ 
        Q2 & Did you feel you got stuck at some point during the training?  \\ 
        Q3 & How much did you enjoy the training?  \\
        Q4 & Did you feel the training should be more difficult for you?  \\
        Q5 & Did you feel you would like the training to be longer with additional tasks to solve?  \\
        Q6 & Would you like to play more cybersecurity training sessions like this one?  \\
        \hline
    \end{tabular}
\end{table}

\Cref{fig:questionnaireResults} presents answers to questions from the post-training questionnaire listed in~\Cref{tab:postGameAssessmentQuestions} for both Junior Hacker and Knowledge Base training.
The participants reported that tasks of both trainings were appropriately designed so that they have successfully completed the training in time (Q1). The majority of participants of both trainings (70\% in Junior Hacker, 68\% in Knowledge Base) did not get stuck \textit{Much} nor \textit{Very much} during the training (Q2).
The participants of both trainings enjoyed the learning experience (Q3). Junior Hacker training was rated higher than Knowledge Base.
The majority of participants (51\% in Junior Hacker, 63\% in Knowledge Base) felt the trainings should be only \textit{Slightly} or \textit{Not at all} more difficult (Q4), which indicates the provided tasks are not overwhelming yet keep the participants appropriately motivated. Only one participant of Junior Hacker training thought the training should be \textit{Very much} more difficult. Next, the participants engaged in both trainings and would like to continue if possible (Q5).
Finally, the participants of both trainings would like to join another similar training (Q6). This was unequivocal for those who participated in Junior Hacker training. Opinions of participants of Knowledge Base training were mixed, though still mainly positive.

To conclude, we see KYPO SLE 
caters to the students with various proficiency. Otherwise, these students would likely have not completed the training if using other state-of-the-art cybersecurity training platforms. 

\subsection{Adaptive Training without Instructor's Supervision}

Since the training sessions with the supervision were a success, we investigated the limits of the proposed approach. We prepared one training session with Knowledge Base training open for two months for anyone interested. 
We expected that the adaptive training would reduce the participants' failure rate and enable them to complete the training as we have seen in our supervised training sessions. However, out of 28 participants joining this session, only 15 successfully finished it. We, therefore, asked these participants who did not have the opportunity to fill in the post-training questionnaire to give us feedback on the training. We specifically asked if the participants encountered any issues during the training. 
Four students provided us with the following  answers:
\begin{enumerate}
    \item \enquote{\textit{I could not for some reason access a file specified by [the] task -- it looked it was not there for some reason, but maybe I did something wrong.}} 
    \item \enquote{\textit{I did not know how to finish the task even with the provided solution.}}
    \item \enquote{\textit{I only started the training to see what is it about. I wanted to play it later, but due to COVID, I didn't manage to do so. I'll try it later.}}
    \item \enquote{\textit{Something interrupted me while participating in this training. Otherwise I would [have] finished [the] whole training.}}
\end{enumerate}

The first two answers may indicate an issue in the design of this particular training, which discouraged the student from continuing. Students tend to stop the training and never come back in such cases.
The third and fourth answer shows these students were forced to stop the training due to unforeseen circumstances that might be more distracting during the unsupervised training.
To conclude, this particular training does not seem suitable for running in the unsupervised mode.

\subsection{Effort Required to Run Adaptive Trainings}

\begin{table}[t]
    \centering
    \caption{Descriptive statistics of training actions and commands entered in KYPO SLE by 86 students (65 from Junior Hacker and 21 from Knowledge Base training). The means are rounded to the nearest whole number.}
    \label{tab:summarizedStatisticsForActionsAndCommands}
    \begin{tabular}{l|r|r|r|r|r}
    \toprule
        Training & Min & Max & Mean & Median & Total \\ \midrule
        \multicolumn{6}{c}{\textbf{Training actions}} \\ \midrule
        Junior Hacker & 7 & 45 & 28 & 25 & 1415 \\  \midrule
        Knowledge Base & 23 & 88 & 43 & 41 & 897 \\  \midrule
        Both & 7 & 88 & 36 & 33 & 2312 \\  \midrule
        \multicolumn{6}{c}{\textbf{Commands}} \\ \midrule
        Junior Hacker & 12 & 155 & 83 & 74 & 4557 \\  \midrule
        Knowledge Base & 54 & 556 & 180 & 150 & 3775 \\  \midrule
        Both & 12 & 556 & 131 & 112 & 8332 \\
    \bottomrule
    \end{tabular}
\end{table}

\Cref{tab:summarizedStatisticsForActionsAndCommands} shows descriptive statistics of training actions and shell commands entered by students who finished the supervised training (86 students, 2 trainings).
Each participant performed 36 actions and typed 131 commands on average during one training session lasting about two hours. In addition, they also filled in the pre-training assessment comprising eight questions. The total amount of data is so vast that it is infeasible to process manually, thus necessitating automation.

To support this argument, we now estimate how much time an instructor familiar with a state-of-the-art environment collecting these data would need to analyze the data manually. Our estimates come from the manual analysis performed in our initial study~\cite{my-2021-FIE-adaptive}.
Without the SLE, the instructor would evaluate the pre-training assessment answers and map them to the relevant training phase. This evaluation may take tens of seconds for each student. Before each training phase, the instructor would need to analyze captured shell commands (searching for keywords, counting the commands) and training actions of each participant (counting the number of wrong answers, searching whether a solution was taken). This analysis may take tens of seconds, perhaps a minute or more in training events with tens of participants or more. This time estimation is based on the experience of four instructors that organized the first four training sessions in~\Cref{table:study-participants} when the SLE was not fully integrated into the KYPO CRP. Finally, the instructor would need to combine all these results to compute the suitable task for each participant using the tutor model. While the instructor is extremely busy and overwhelmed at that time, the student is only waiting to be assigned the next task. Using this \enquote{manual} approach, the instructor can handle only a few students. However, for medium to large classes, the manual approach does not scale. This example clearly supports the necessity of a SLE for running adaptive hands-on cybersecurity training sessions.
What is more, automated task assignments by the SLE enable instructors to focus on providing additional help to struggling students.

\subsection{Limitations}

In this evaluation, the Knowledge Base training has only two tasks in each phase. Providing more tasks may increase the probability that the participant will get a more suitable task and increase their overall student experience.

We challenged our approach and studied whether the SLE can fully substitute a human instructor. The results of Knowledge Base training in an unsupervised mode showed this is still not feasible. However, we might obtain better results with the Junior Hacker training, which we consider easier than Knowledge Base.

Another aspect that may negatively affect the unsupervised training session is that the SLE cannot easily recognize whether the student is thinking about the task (while not producing any training action or typing the command) or interrupted the training for a while. The latter may mislead the tutor model using the \enquote{completed in time} metric.

\subsection{Lessons Learned}\label{subsec:lessons_learned}
For easier adoption of the developed SLE, we highlight the main lessons learned and provide general recommendations. All lessons are based on our experience from adaptive trainings in an authentic setting. Each lesson is illustrated with a concrete example.

\subsubsection{Adjust the weights in the model carefully} Inappropriate settings of weights in the decision matrices of the tutor model may lead to suboptimal transitions through the training tasks. The instructor(s) should verify the training with simulated students who perform differently to test that the model weights are set correctly. To reduce the complexity of such simulation, the instructor can use assisting tools described in~\cite{seda2022adaptive}.

Next, the instructor may stress critical prerequisites for a particular phase by setting a greater value of an important weight. For instance, all weights but one were set to one in the Knowledge Base training. The weight of timely completion in Phase 4 was set to two for Phase 7 to express its importance.

\subsubsection{The training content must be thoroughly designed and tested} The SLE significantly helps the instructor to prepare and run the adaptive hands-on cybersecurity training. However, when the training content is not designed properly, (e.g., long and difficult Phase 6 in the Knowledge Base training), the students might get stuck in the task due to the misunderstanding of the task or the insufficient number of easier tasks. To design trainings more effectively, instructors may benefit from the documented guidelines~\cite{2021-SIGCSE-poster-guidelines}.

\subsubsection{The beginning of the training affects its progress} The training sessions are mostly held in a limited time frame (such as class). 
The pre-training assessment questionnaire should be brief and follow best practices for educational assessment~\cite{astin2012assessment, petty2009}. It should be also complemented by one or two phases with a single task that evaluates the skills of the students. For instance, Phase 1 in the Knowledge Base training served this purpose. The combination of quizzes, skill self-assessment, and skill evaluation provides a solid foundation for the tutor model.

\subsubsection{Design as many tasks for each phase as possible} To cater to students of various proficiency, the training should provide several variant tasks in each phase. If there are only two tasks in a phase as in the Knowledge Base training, some students may still struggle and need the instructor's assistance. However, a higher number of tasks increases the instructor’s effort in preparing the training.

\subsubsection{Design at least some relationships between the training phases} KYPO SLE relies on the collected data and the model settings. If the instructor sets the model weights so that there are no relationships between any phases, tasks will be assigned only based on the pre-training assessment questionnaire. This might not truly reflect the students' proficiency before entering particular tasks.

\section{Conclusion}
\label{sec:conclusion}

The proposed smart learning environment KYPO SLE is, to the best of our knowledge, one of the first SLEs for hands-on cybersecurity training. The main objective of KYPO SLE is to provide an optimal individual learning path in hands-on training to improve the students' experience. To achieve that, we designed a new tutor model and a new training format that supports a graph structure to enable different learning paths for each student. The tutor model processes questionnaire answers and training actions from the learning management system and shell commands from the emulated environment. Based on these data, it determines the most suitable task for each individual in the training.

We implemented the training format, data collection, and the tutor model and evaluated the developed SLE with 114 participants from a wide variety of institutions (high schools, universities, and companies). The evaluation showed that the proposed tutor model and adaptive training format are generic enough to be used for various training sessions with different topics. Further, the developed SLE can increase the students' ability to successfully complete the hands-on training, and thus increase their positive experience. Without the SLE, instructors would not be able to process the complex and voluminous learning data required for determining the most suitable task. Finally, to ease the adoption of the proposed SLE, we released it as an open-source project~\cite{kypo-website} together with a detailed documentation~\cite{kypo-website-documentation} and an exemplary definition of an adaptive training~\cite{junior-hacker-adaptive}.

\subsection{Affordances of KYPO SLE}

Our smart lab qualifies as a SLE because it fulfills the six characteristic features identified by Tabuenca et al.~\cite{Tabuenca2021}. Specifically, it is or has:
\begin{itemize}
    \item \emph{Adaptable} -- it adjusts the learning environment so that it is adaptive and personalized for each student.
    \item \emph{Tracking and monitoring} -- the instructor can monitor progress of each student during the training and revisit the results of each individual student after the training.
    \item \emph{Feedback and recommendations} -- tasks assigned to students are determined based on the student's assessment and current performance,
    \item \emph{Pattern recognition} -- the instructor can define patterns  that are searched for in students' data during the training. These patterns are essential for selecting the most suitable task for each student.
    \item \emph{Efficient} -- the lab enables assigning tasks of appropriate difficulty with respect to students' proficiency and current performance.
    \item \emph{Effective} -- the lab enables more students to complete the training compared to the non-adaptive training where all students are provided with the same tasks regardless of students' proficiency and performance.
\end{itemize}

\subsection{Open Challenges}

We identified two distinct directions for possible future work. 

\subsubsection*{Machine learning for setting the tutor model}
The parameters of the tutor model are now set by instructors based on their expertise, the content of the tasks, and their relations between phases. Exploring how to employ machine learning algorithms should optimize metrics selection and weights settings. The application of machine learning algorithms will be challenging due to the typically small number of participants in each training session, their diverse proficiency, and the complexity of performed tasks.

\subsubsection*{Conditional phases}
The current format of the adaptive training assumes each student will pass through each training phase. Enhancing the format by allowing to skip some phases if certain conditions are met during the training can open new opportunities.

\section*{Acknowledgment}

This research was supported by the ERDF project \textit{CyberSecurity, CyberCrime and Critical Information Infrastructures Center of Excellence} (No. CZ.02.1.01/0.0/0.0/16\_019/0000822).
The authors thank all researchers and developers of KYPO Cyber Range Platform who transferred research ideas into real open-source software. 
This paper is an expanded version of a paper that was published in the proceedings of 2021 IEEE Frontiers in Education Conference (FIE).

\bibliographystyle{IEEEtran}
\bibliography{IEEEabrv.bib, references.bib}

\newcommand{\upshift}{\vspace*{-1.5cm}}

\begin{IEEEbiography}[{\includegraphics[width=1in,height=1.25in,clip,keepaspectratio]{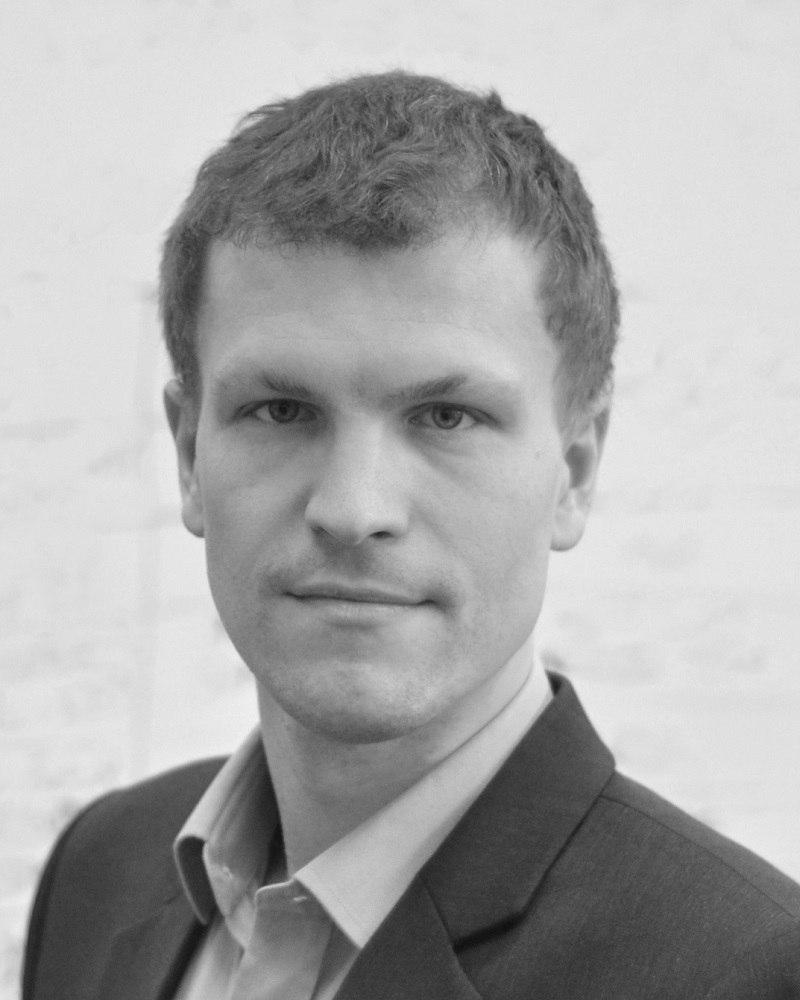}}]{Jan Vykopal}
is an assistant professor with Masaryk University, Brno, Czech Republic. He teaches cybersecurity and researches how to teach it better. He has been with KYPO Cyber Range Platform since its early beginnings in 2013. He has been designing and organizing various cybersecurity games and exercises, including the Czech national defense exercise, since 2015. Jan also organizes summer schools for finalists of the Czech national cybersecurity competition.
\end{IEEEbiography}
\upshift

\begin{IEEEbiography}[{\includegraphics[width=1in,height=1.25in,clip,keepaspectratio]{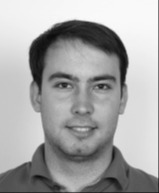}}]{Pavel Seda} has defended his Ph.D. thesis in electrical engineering at Brno University of Technology in 2022. Further, he received his MSc. degree in communications and informatics from the Brno University of Technology and MSc. in Applied Informatics from Masaryk University. From 2014 to 2018 Pavel worked as Java Developer at IBM. Currently, he focuses on research topics on cybersecurity, technologies, optimization, and cybersecurity education that are investigated in research projects that Pavel is involved.
\end{IEEEbiography}
\upshift

\begin{IEEEbiography}[{\includegraphics[width=1in,height=1.25in,clip,keepaspectratio]{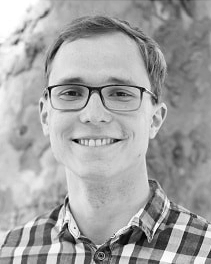}}]{Valdemar Švábenský}
received the Ph.D. degree in data-driven support of hands-on cybersecurity training from Masaryk University in 2022. His research interests include methods for automatic analysis of training data to generate tailored feedback for students and instructors. Dr. Švábenský received the Best Paper Award at the ACM SIGCSE Technical Symposium 2020 and 2022. He also received two university-wide awards for the contribution to teaching computer science.
\end{IEEEbiography}
\upshift

\begin{IEEEbiography}[{\includegraphics[width=1in,height=1.25in,clip,keepaspectratio]{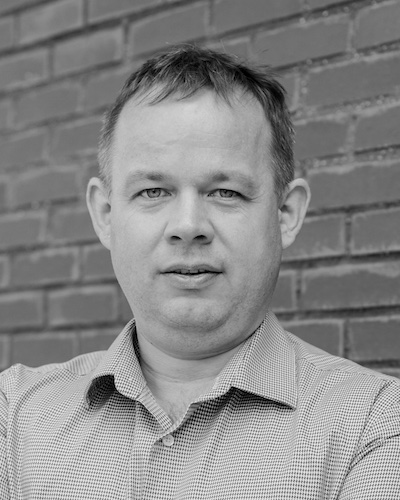}}]{Pavel Čeleda}
is an associate professor at Masaryk University. He received a Ph.D. in Informatics from the University of Defence, Brno. His main research interests include traffic analysis, situational awareness, and cybersecurity testbeds for research and education. These research topics are the subject of many projects, collaborations, and supervised Ph.D. theses. He is a principal investigator of the KYPO Cyber Range project and co-principal investigator of the C4e Center of Excellence.
\end{IEEEbiography}





\end{document}